\documentstyle[12pt]{article}

\begin{document}



\topmargin=-.2in
\textheight=8.2in
\textwidth=6.0in

\def\beq{\begin{equation}}
\def\eq{\end{equation}}
\def\eeq{\end{equation}}

\newcommand{\gsim}{\lower.7ex\hbox{$\;\stackrel{\textstyle>}{\sim}\;$}}
\newcommand{\lsim}{\lower.7ex\hbox{$\;\stackrel{\textstyle<}{\sim}\;$}}

\begin{titlepage}

\begin{flushright}
SU-ITP-00-21\\
\end{flushright}

\vspace{.3in}

\huge
\begin{center}
 Holographic Vacuum Energy
\end{center}

\large


\vspace{.1in}
\begin{center}

Scott Thomas\\
\vspace{.15in}
Physics Department \\
Stanford University \\
Stanford, CA  94305 \\

\end{center}

\vspace{0.2in}


\normalsize

\begin{abstract}
{Gravitational holography is argued to
render the cosmological constant stable against divergent
quantum corrections, 
thus providing a technically natural
solution to the cosmological constant problem.
Evidence for quantum stability of the cosmological
constant is illustrated in a number of examples including
bulk descriptions in terms of delocalized degrees of freedom,
boundary screen descriptions on stretched horizons, and
non-supersymmetric conformal field theories as dual descriptions
of anti-de Sitter space.
In an expanding universe, holographic
quantum contributions to the stress-energy tensor are argued
to be at most of order the energy density of the dominant matter
component.}
\end{abstract}

\end{titlepage}

\baselineskip=18pt



The cosmological constant problem has been 
the most severe hierarchy problem for fundamental physics.
In a 
local quantum field theory,
the cosmological constant, or vacuum
energy density, 
receives divergent quantum contributions from zero point
energies of virtual field theory degrees of freedom.
The divergence is expected to be controlled by high-energy short distance
physics at an ultraviolet scale $M \sim \hbar / \delta$,
such as the Planck or supersymmetry breaking scale,
$\Lambda_D \sim \hbar / \delta^D$,
where $D$ is the number of space-time dimensions.
The relatively slow expansion of the universe, however, implies
that the cosmological constant is very much smaller,
$\Lambda_4 \lsim {\rm meV}/ (100~\mu{\rm m})^3$.

A complete formulation of quantum gravity with which to address
the cosmological constant problem is not yet available.
However, the Bekenstein bound on the entropy of a space-like
region \cite{bek} follows from the thermodynamic properties
of large black holes \cite{bekbh,hawkbh}, and is
likely to be a feature of any theory;
$S \leq A_{D-2} /(4 G_D \hbar)$, where $G_D$ is the
$D$-dimensional
Newton constant, and $A_{D-2}$ is 
the area bounding
any region which satisfies the space-like projection
theorem \cite{bcov}.
This bound led 't Hooft \cite{thooftholo}
and Susskind \cite{susskindholo} to propose the holographic
principle in which the number of independent degrees of freedom,
equivalent to the maximum entropy, is also subject to the
Bekenstein bound
\beq
N_{\rm dof} \leq A_{D-2} /(4 G_D \hbar)
\label{ndof}
\eq
The number of degrees of freedom 
within a region satisfying the space-like projection
theorem is then finite and extensive in the 
bounding area \cite{thooftholo,susskindholo,bholo} rather than the volume.
A local quantum field theory description of gravity, with
a divergent number of
volume extensive degrees of freedom, is clearly inconsistent
with holography.
The volume divergent local field theory 
estimate given above for the quantum vacuum energy is then unlikely 
to be correct 
since the states required to give this result do not exist in a
holographic theory of gravity. 

In this letter it is argued that holography
renders the cosmological constant stable against
divergent quantum corrections, and allows only finite
corrections, thereby
providing a technically natural solution to the cosmological
constant problem.
As discussed below, 
this follows first from the holographic reduction in the number
of independent degrees of freedom, and second from the holographic 
energy per degree of freedom. 
A natural solution 
can then follow from a symmetry of the action.
Previous discussions of holography have attempted to relate
the cosmological constant to supersymmetry breaking \cite{banks},
to correlated infrared and ultraviolet cutoffs in an effective field
theory \cite{ckn},
and to a probabilistic interpretation 
of large entropy static universes 
\cite{horavaminic}.

The emergence of small scale local physics from a general
holographic description is an open problem and will not be addressed
here.


{\it Holographic Vacuum Energy}---
The local field theory estimate of 
quantum vacuum energy involves 
degrees of freedom with density determined
by the ultraviolet scale $M$.
However, the actual volume density of non-redundant
holographic degrees of freedom
(\ref{ndof}) is determined by an infrared scale related to the
bounding area.
Quantum corrections to the vacuum energy
of a 
low curvature 
space-time background 
from these non-redundant degrees of freedom must therefore
involve an infrared scale.
The magnitude of the corrections depends
not only on the the number of degrees of freedom but also the 
quantum energy per degree of freedom.
Parametric estimates employing non-redundant degrees of freedom
for a number of holographic descriptions including 
non-local bulk descriptions, virtual black holes, holographic
screens, and conformal field theory descriptions of anti-de Sitter
spaces are detailed below.
In all cases holographic quantum corrections
to the vacuum energy density result in only a finite
shift of the classical value, 
and parametrically do not destabilize the background geometry.


{\it Bulk Holography}---
Any holographic description of gravity
in terms of local bulk fields should be invested with 
a holographic gauge symmetry \cite{hgs}
which eliminates redundant
volume extensive degrees of freedom.
An explicit 
Faddeev-Popov procedure for eliminating redundant 
degrees of freedom in a general setting 
is not yet available.
However, gauge invariant or gauge fixed
descriptions may exist in which the non-redundant
degrees of freedom are manifest.
For a generic 
background, the largest space-like
regions which obey the holographic bound (\ref{ndof}) are generally
of order the background curvature scale
[the special case of anti-de Sitter space in which
all regions satisfy the space-like projection theorem
is discussed below].
In a gauge appropriate to calculating a global quantum effect on the
background geometry,
it is then natural to postulate that
uniformly volume distributed
bulk holographic degrees of freedom are delocalized
on the scale of the background radius of curvature, $R$, since this
is the relevant holographic length scale.

The Heisenberg quantum energy of each delocalized holographic
degree of freedom is $E_{\rm dof} \sim \hbar /R$.
The quantum contribution to the global vacuum energy density,
$\delta \Lambda_D \sim N_{\rm dof} E_{\rm dof} / R^{D-1}$,
with $A_{D-2} \sim R^{D-2}$, is then
\beq
\delta \Lambda_D \lsim 1/ (G_D R^2)
\label{lambdashift}
\eq
Since the factors of $\hbar$ cancel in the product 
$N_{\rm dof} E_{\rm dof}$, this shift should
appear as a classical bulk effect.
Einstein's equation for the classical background
relates the background energy density and radius of curvature by
$ \rho \sim 1 /(G_D R^2)$. 
So holographic quantum contributions to the vacuum energy
are parametrically at most of order the background energy density.
Note that the postulate of delocalized bulk holographic degrees of
freedom implies that quantum contributions to the vacuum
energy satisfy the gravitational mass bound
$ N_{\rm dof} E_{\rm dof} \lsim R^{D-3}/ G_D$.

It has been suggested that quantum contributions to certain
quantities, specifically in de Sitter space, are
dominated by large virtual black holes at the
curvature scale of the background geometry \cite{banks}.
This may occur in thermodynamic averages which are entropy dominated
by large black holes, or because virtual
contributions at super-Planckian energies are dominated by
massive black holes \cite{banks}.
If a gauge fixing of the bulk holographic gauge symmetry
does in fact exist in which this is an appropriate alternate description,
it is clear that large virtual black holes at the background curvature
scale, 
with entropy that saturates the bound (\ref{ndof}),
energy $E \sim R^{D-3}/ G_D$, and volume $R^{D-1}$,
contribute to the vacuum energy density parametrically by the same
estimate (\ref{lambdashift}) given above.


{\it Holographic Screens}---
Since the number of holographic degrees of freedom
in regions satisfying the space-like projection theorem are
limited by the bounding area, in many circumstances there
may be a gauge fixing of the holographic gauge symmetry
in which the non-redundant degrees of freedom
can be localized on $D-2$ dimensional holographic
screens \cite{thooftholo,susskindholo}.

For definiteness consider the case of an expanding spatially flat
Friedman-Robertson-Walker cosmology with
Hubble parameter $H$ and
line element
\beq
ds^2 = (1 - r^2 H^2) ~dt^2 + 2rH ~dr dt - dr^2 - r^2 ~d \Omega_{D-2}^2
\label{dsFRW}
\eq
The apparent horizon, at a distance $r=1/H$ from any observer,
forms a preferred screen which
encodes the entire space-time \cite{bholo} or the normal
space-like region within the horizon.
The holographic bound (\ref{ndof}) implies that the area density
of non-redundant degrees of freedom
on the screen is limited by the Planck density,
$N_{\rm dof}/A_{D-2} \leq 1/ (4 \ell_P^{D-2})$
where $\ell_P = (G_D \hbar)^{1/(D-2)}$.
The Heisenberg quantum 
energy of each screen degree of freedom is then limited by
$E_{\rm dof} \lsim \hbar / \ell_P$, and the total
quantum energy of the screen is limited by
$N_{\rm dof} E_{\rm dof} \lsim R^{D-2} / (G_D \ell_P)$, where here
$R = 1/H$.
The precise procedure for projection of screen energy into the bulk is
not known, but parametrically should involve a
screen--bulk redshift factor.
The apparent horizon is a surface of infinite redshift, so
a regulated screen 
must be employed.
The area density and energy limits above imply that the
Planck length and energy may be taken as limiting
cutoffs for local screen degrees of freedom.
So a stretched screen one Planck distance inside the apparent
horizon in the screen frame
is appropriate,
in analogy with the stretched horizon of a black hole \cite{stretched}.
The stretched screen--bulk inverse redshift factor,
$\sqrt{g_{00}} \simeq \sqrt{2} \ell_P/R$, then
gives a bulk quantum energy of $E \lsim R^{D-3} / G_D$.
Averaged over the enclosed volume, $R^{D-1}$,
this gives a shift of the vacuum energy density
identical to the estimate (\ref{lambdashift}) given above.
This 
applies to other space-times with apparent horizon preferred screens obtained
by space-like projection \cite{bholo}
such as de Sitter space.


{\it Conformal Field Theory Description of
Anti-de Sitter Space}---
The best understood example of holography is the duality between
$D$-dimensional anti-de Sitter space 
and a $D-1$ dimensional
boundary conformal field theory \cite{maldacena}.
The description of 
IIB superstring theory or supergravity in an $AdS_5 \times S^5$
background, or the consistent truncation to 
${\cal N}=8$ gauged supergravity in $AdS_5$, in terms of an 
${\cal N}=4$ supersymmetric
$SU(N)$ boundary gauge theory with both large $N$ and $g^2N$
provides a number of non-trivial
checks of this duality \cite{adscftreview}.
This includes an infrared--ultraviolet
duality between the bulk and boundary descriptions
which implies a holographic bound (\ref{ndof})
on lattice regulated conformal field theory degrees of
freedom within space-like regions large compared to the
anti-de Sitter radius \cite{susskindwitten}.

The dual conformal field theory description of anti-de Sitter space
may be thought of as
a gauge fixing of the presumed holographic gauge symmetry
in terms of
non-redundant 
degrees of freedom.
The conformal field theory energy-momentum tensor
acts as a source for the bulk metric.
Quantum 
vacuum energy 
could be
addressed by a duality with a non-supersymmetric boundary
conformal field theory, in which zero point energies do not
cancel.
A purely non-supersymmetric example of this type is not yet know,
but should exist
if large non-supersymmetric anti-de Sitter spaces are
consistent backgrounds in a quantum theory of gravity.

Quantum contributions to 
anti de-Sitter space vacuum energy
can be estimated parametrically from the holographic properties of the
duality.
The 
anti-de Sitter line element near the
asymptotic boundary, $y \rightarrow 0$, is
\beq
ds^2 \simeq (R^2/y^2)(dt^2 - dy^2 - R^2~ d \Omega_{D-2}^2 )
\eq
with
$\Lambda_D = -(D-1)(D-2)/(16 \pi G_D R^2)$.
Spatial sections of the boundary theory are defined on
$S^{D-2}$ with radius $R$.
In order to regulate the divergent vacuum energy of a non-supersymmetric
boundary 
theory, it is convenient to employ
block spin renormalization in which the theory is
defined on an infinite family of spatial lattices
labeled by a renormalization step 
$n \in {\rm {\bf Z}}$,
with lattice spacings related by
$\delta_{n+1} = \delta_n/2$.
The 
Heisenberg energy per lattice degree of freedom
is $E_{\rm dof} \sim \hbar /  \delta_n$.
The boundary theory vacuum energy density
at renormalization step $n$ is then
$\Lambda_{D-1;n} \sim N_{\rm dof} E_{\rm dof} / \delta_n^{D-2} \sim
\hbar N_{\rm dof} / \delta_n^{D-1}$ where here
$N_{\rm dof} \sim c$ is the effective
number of boundary degrees of freedom
per lattice site, and $c$ is the central charge. 
The boundary theory vacuum energy density 
is ultraviolet divergent for $n \rightarrow \infty$.

Scale--radius duality implies that
a lattice scale $\delta_n$ corresponds
to a radial $y$ coordinate of $y \sim \delta_n$ \cite{susskindwitten}.
The bulk proper area covered by a lattice at renormalization step $n$ 
is then 
$A_{D-2;n} \sim ( \sqrt{h_{ii}} R)^{D-2} \sim (R^2/ \delta_n)^{D-2}$,
where $h_{ii} \simeq R^2/y^2$ is the bulk transverse spatial metric
for $y \sim \delta_n \ll R$. 
Lattices which differ by one renormalization step are 
separated by a proper bulk radial distance of
$\Delta r \simeq   \int_{\delta_n}^{\delta_n/2} \! \sqrt{g_{yy}}~dy =(\ln 2) R$.
The contribution to the bulk energy density of any given 
lattice 
is then spread over this radial distance, and the bulk volume 
covered by this lattice is 
$V_n \sim A_{D-1;n} \Delta r \sim (R^2 / \delta_n)^{D-2} R$.
Now the boundary total quantum energy of the lattice at renormalization 
step $n$ is $\tilde{E}_n \sim \Lambda_{D-1;n} R^{D-2}$, and the 
associated proper bulk energy is 
$E_n \sim \tilde{E}_n / \sqrt{g_{00}} \sim \tilde{E}_n \delta_n /R$ where
$g_{00} \simeq R^2/y^2$. 
So the quantum shift of
the proper bulk cosmological constant $\delta \Lambda_D \sim E_n / V_n$ is
\beq
\delta \Lambda_D \sim 
\Lambda_{D-1;n} /
\left( \sqrt{g_{00}} ( \sqrt{h_{ii}} )^{D-2} R \right)
\sim ~\hbar N_{\rm dof} / R^D
\label{AdSshift}
\eq
independent of $n$. 
The boundary theory divergent $\Lambda_{D-1}$
is therefore distributed along
the bulk radial direction as a constant finite shift,
$\delta \Lambda_{D}$,
of the bulk cosmological constant.
Any consistent duality 
must satisfy the holographic bound (\ref{ndof}),
which implies that the shift in bulk cosmological constant
(\ref{AdSshift}) is
identical to the estimate (\ref{lambdashift}) given above.
Note that since the bulk proper lattice spacing and Heisenberg
energy per degree of freedom are both determined by
$R$, the boundary 
theory degrees of freedom
are effectively delocalized on the scale of the background
curvature, as conjectured above for the general case of bulk
holographic degrees of freedom.

Certain non-supersymmetric
boundary theories can be 
obtained by a relevant perturbation of an ultraviolet
supersymmetric conformal field theory \cite{nonSUSYboundary}.
In anti-de Sitter space the relevant operator
corresponds to a closed $S^{D-2}$
domain wall which separates an exterior region
associated to the ultraviolet supersymmetric theory from an
interior region associated to the infrared non-supersymmetric theory.
In this case the divergent
quantum vacuum energy of the infrared boundary theory, $\Lambda_{D-1}$,
corresponds to a finite classical
auxiliary expectation value which breaks supersymmetry
in the interior anti-de Sitter region \cite{toappear}.
The relevant operator cuts off the divergence in the boundary
theory, and the auxiliary expectation value 
vanishes in the
exterior anti-de Sitter region.

For any duality of the type discussed above,
the central charge of the
boundary conformal field theory
is related, at least for $D$ odd,
to the anti-de Sitter space radius.
For definiteness consider $D=5$ for which
$R = (8c  G_5 \hbar / \pi)^{1/3}$ \cite{AdSc} or
$\Lambda_5 =-(3/4) / ( 64 \pi c^2  G_5^5 \hbar^2)^{1/3}$.
In this case
the boundary central charge, 
and therefore the bulk cosmological constant,
is modified with respect to the
classical free field value by quantum anomalous dimensions
of the boundary fields.
In this language a finite shift $\delta \Lambda_5$ due to boundary
quantum effects is qualitatively independent of whether the
boundary conformal field theory is supersymmetric with vanishing boundary
vacuum energy, $\Lambda_{4}=0$, or non-supersymmetric
with divergent $\Lambda_{4}$.


{\it The Cosmological Constant Problem }---
Any hierarchy problem
has two aspects.
The first is technical naturalness associated with 
stability of the small or vanishing parameter characterizing the
hierarchy.
This generally follows from a symmetry principle which
forbids naive 
corrections that could destabilize
the hierarchy.
In all the holographic descriptions given above,
the implicit holographic gauge symmetry is responsible for 
quantum stability of the 
vacuum energy since it
removes redundant volume extensive degrees of freedom that naively would
give an ultraviolet divergent contribution.
Both the number and Heisenberg energy
of the non-redundant holographic degrees of freedom are controlled
by the infrared scale set by the background curvature.
This 
should arise as a general feature in
any holographic theory of gravity.
Even though the cosmological constant is a relevant operator
from the local bulk point of view,
holographic gauge symmetry protects it from divergent 
corrections, 
and renders it stable against small changes in ultraviolet 
parameters. 
Holography 
provides a technically natural solution
to the cosmological constant problem.

The second aspect of any hierarchy problem, which may be addressed
once a technically natural solution is understood, is the
natural value of the small or vanishing parameter.
With any technically natural hierarchy
it is possible
simply to take the small
parameter as input, without
specifying an underlying natural mechanism for its value.
In the present context this amounts to 
taking the number of holographic degrees of freedom 
as input, at least for static universes. 

A naturally vanishing cosmological constant could follow
from an exact symmetry of the full theory.
As long as the symmetry commutes with holographic gauge symmetry
the local bulk action should be invariant,
even without implementing a Faddeev-Popov procedure to
remove redundant degrees of freedom.
The local scalar potential
in general relativity may be written as
\beq
V(\phi) = 
(D-2)~D_iW g^{ij} D_jW -  8 \pi G_D(D-1)~W^2
\label{vphi}
\eq
where
$W=W(\phi_i)$ is an arbitrary pre-potential function,
$g^{ij}$ is the inverse field space metric, and
$D_i$ is a field space covariant derivative.
The general form (\ref{vphi}) follows from
perturbative stability of anti-de Sitter space
solutions \cite{boucher}, exact stability with a single
scalar \cite{townsend},
or directly from Einstein's equations in certain ansatz.
If $V$ is invariant but 
$W$ transforms under an exact symmetry, and
all the scalars transform under additional exact
symmetries,
Schur's Lemma ensures that both $D_i W(0)=0$ and
$W(0)=0$.
The symmetries may be continuous or discrete,
either global or gauged.
The origin of field space, $\phi_i=0$, is then
an enhanced symmetry
extremum with vanishing potential, $V(0)=0$.
Given that Lorentz scalars which break electroweak and quark
chiral symmetries are non-zero in the vacuum,
this option 
may not be applicable to our universe.

Another possibility for a naturally vanishing cosmological
constant is the existence of an exact symmetry
which enforces a relation between $D_iW$ and $W$
in (\ref{vphi}) such that $V=0$.
This is applicable to supergravity and superstring
theories in $D=4$ with spontaneously broken ${\cal N}=1$
supersymmetry which requires $D_iW \neq 0$.
Classical non-compact symmetries of this type exist in
no-scale supergravity models \cite{noscale}.
Dimensional reduction of 
superstring theories
with a non-vanishing superpotential from, for example,
race-track stabilized gaugino condensation,
can give rise to no-scale symmetries for moduli either from
a classical scale invariance \cite{noscaleinv}
or inheritance from extended supersymmetries of the underlying
theory.
These symmetries are generically broken in the
four dimensional quantum local potential,
but might be recovered at 
points on moduli space.

Another possibility is that
a manifestly holographic or boundary description
has an instability which is spontaneously 
stabilized by a finite number of holographic degrees
of freedom in the ground state.
If the number of spontaneous degrees of freedom is divergent, then flat
space appears naturally.

For any input or natural explanation of the cosmological constant hierarchy,
holography renders the cosmological constant stable and
a small or vanishing value technically natural.


{\it Cosmology}---
Holographic 
contributions to the vacuum energy
can have important implications for cosmology.
In all the holographic descriptions given above, the
quantum vacuum energy density is parametrically limited to be
at most of order the background energy density.
In an expanding universe this implies that quantum 
corrections
to the stress-energy tensor at any epoch are at most of order the
energy density of the dominant matter component at that epoch \cite{toappear}.
This is natural since the infrared holographic length scale
is determined by the holographic screen
on the apparent horizon.



{\it Conclusions}---
The holographic interpretation \cite{thooftholo,susskindholo}
of the Bekenstein bound \cite{bek} implies that the number of
degrees of freedom and Hilbert space dimensionality
of a gravitating system is finite.
Local field theory, even for example (super)string or M-theory
interpreted as a local field theory below the Planck
or string scale,
can not be a fundamental description,
and is likely to 
break down for processes which nearly saturate the
Bekenstein bound.
This is certainly true for real processes which involve
excited degrees of freedom, 
but as advocated
here, should also be true for virtual effects such as
quantum contributions to the vacuum energy or equation of
state of an expanding universe.
It is precisely for such effects that holographic gauge
symmetry, absent in a bulk local field theory description, becomes
important since these effects necessarily involve all the available
non-redundant virtual degrees of freedom. 
The most challenging problem for a holographic theory of gravity
may not be the cosmological constant problem,
but rather how local small scale physics arises from a holographic
description.


I would like to
thank M. Dine, L. Dixon, M. Peskin,
S. Shenker, and L. Susskind for useful discussions.
This work was supported by
the US National Science Foundation under grant PHY98-70115,
the Alfred P. Sloan Foundation, and Stanford
University through the Frederick E.~Terman Fellowship.



\begin{thebibliography}{99}
\bibitem{finitenote} Fully consistent finite quantum theories
of gravity in 
$AdS_5 \times {\cal M}$, where ${\cal M}$ is a compact manifold,
may only be dual to 
finite conformal field theories in four dimensions.
\end{thebibliography}
\end{document}